\documentclass[epsf]{mn2e}
\usepackage{graphicx}
\usepackage{times}
\def\mmm{(m-M)$_0$}
\def\ebv{$E(B-V)$}
\def\evi{$E(V-I)$}

\def\msun{M$_{\odot}$}
\def\gsim{\;\lower.6ex\hbox{$\sim$}\kern-7.75pt\raise.65ex\hbox{$>$}\;}
\def\lsim{\;\lower.6ex\hbox{$\sim$}\kern-7.75pt\raise.65ex\hbox{$<$}\;}

\title[NGC 3960]{Photometric and spectroscopic study of the intermediate age 
open cluster NGC 3960\thanks{This work is based on observations collected at the European Southern
Observatory, Chile (Programs 66.A-0485, and 67.D-0014). }
}

\author[Bragaglia et al.]{A. Bragaglia$^1$, M. Tosi$^1$, E. Carretta$^1$, 
 R.G. Gratton$^2$, G. Marconi$^3$, E. Pompei$^3$
 \\
 \\
$^1$ INAF--Osservatorio Astronomico di Bologna, Via Ranzani 1, I-40127 Bologna,
      Italy, 
      e-mail angela.bragaglia, monica.tosi @bo.astro.it \\
$^2$ INAF--Osservatorio Astronomico di Padova, vicolo Osservatorio 5, I-35122
Padova, email gratton@pd.astro.it  \\
$^3$ ESO, Alonso de Cordova 3107, Vitacura, Santiago, Chile,
e-mail gmarconi, epompei @eso.org}

\date{}

\begin{document}
\maketitle

\begin{abstract}
We present CCD $UBVI$ photometry and high-resolution spectroscopy of the 
intermediate age open cluster NGC 3960. 
The colour - magnitude diagrams (CMDs)
derived from the photometric data and interpreted with the synthetic CMD
method allow us to estimate the cluster parameters. We derive: 
age $\tau =$ 0.9 or 0.6 Gyr (depending on whether or not overshooting from
convective regions is included in the adopted stellar models), distance 
\mmm = 11.6 $\pm$ 0.1, reddening \ebv = 0.29 $\pm$ 0.02, differential reddening
$\Delta$\ebv = 0.05 and approximate metallicity between solar and half of
solar. 
We obtained high resolution spectra of three clump stars, and derived an
average [Fe/H] = $-0.12$ (rms 0.04 dex), in very good agreement with the
photometric determination. We also obtained abundances of $\alpha$-elements,
Fe-peak elements, and of Ba. 
The reddenings toward individual stars derived from the spectroscopic temperatures and
the Alonso et al. calibrations give further support to the existence of
significative variations across the cluster.
\end{abstract}

\begin{keywords}
Hertzsprung-Russell (HR) diagram -- open clusters and associations: general --
open clusters and associations: individual: NGC 3960 -- stars: abundances
\end{keywords}

\begin{figure*}
\begin{center}
\includegraphics[bb=36 285 576 510, clip, scale=0.93]{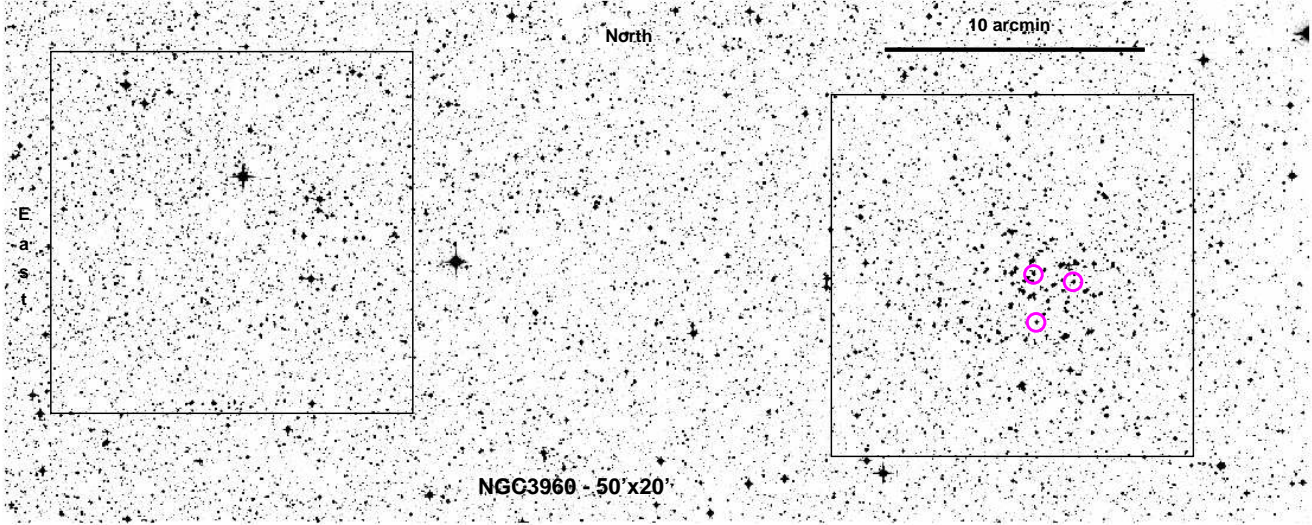}
\caption{Map of the observed fields: the right-hand square indicates the 
pointing on the cluster centre, while the left-hand one indicates the external
comparison field (each one of 13.5 $\times$ 13.5 arcmin$^2$). The three circles indicate
the stars observed with FEROS.}
\label{fig-map}
\end{center}
\end{figure*}

\section{Introduction}

Open clusters (OCs) have been recognized since a long time to be among the
best objects to study the Galactic disk (e.g., Janes \& Phelps 1994, Friel
1995, Twarog, Ashman \& Anthony-Twarog  1997). We have an on going project
aimed at exploiting their ability to trace the present properties of the disk,
its history and evolution by collecting photometric and spectroscopic data of a
large sample of old OCs and deriving in a precise and homogeneous way their
fundamental properties (age, distance, reddening, chemical abundance).
An updated review of our general project, although mostly focussed on the
photometric part, can be found in Bragaglia \& Tosi (2005). 

This paper is devoted to NGC 3960 (C 1148-554), an intermediate age open cluster
located at RA(2000) = 11:50:40, DEC(2000) = --55:40:28,  or l = 294.41,  b = 6.18.
We have acquired CCD photometry of two fields and high-resolution spectra of three
stars, which have allowed us to  derive  the cluster age, distance, reddening and
chemical abundances.

There are two main literature papers about this cluster based on photometry.
Janes (1981, hereafter J81) first published its CMD,  based on BV photographic
photometry for 318 stars, 98 of which have also photoelectric measures.
His estimate of the cluster parameters are: \mmm = 11.1 $\pm$ 0.2, E(B-V) =
0.29 $\pm$ 0.02, and age slightly older than the Hyades.
More recently, Prisinzano et al. (2004, hereafter P04) used data obtained
with the Wide Field Imager at the ESO-Max Planck 2.2m telescope in the $B, V, I$
bands to determine an age between 0.9 and 1.4 Gyr and \mmm = 11.35 using
isochrones with Z = 0.01, the metallicity given by Friel et al. (2002). They
also inferred the luminosity (and mass) function. They find strong
indications of differential reddening, with \ebv \ varying from 0.16 to 0.62 over
the about 30 $\times$ 30 arcmin$^2$ field of view, 
with a value of 0.29 near the cluster centre.  

The metallicity of NGC 3960 has been found to be subsolar by many studies, but
the individual estimates range from [Fe/H] = $-$0.68 to $-$0.06. 
J81 obtained DDO data for six giants, and derived [Fe/H] = $-$0.3 $\pm$
0.06, while Flynn \& Mermilliod (1991) and Piatti, Clari\`a \& Abadi (1992),
using improved DDO abundance calibrations, derived [Fe/H] =
$-0.19$, and $-0.06$, respectively. 
Friel \& Janes (1993), using low resolution spectroscopy, found [Fe/H] =
$-$0.34 $\pm$ 0.08, and determined the membership status of 7 giants. 
Geisler, Clari\'a \& Minniti (1992), using  Washington photometry
of the  cluster giants, determined [Fe/H] = $-$0.68 $\pm$ 0.28.
Paunzen \& Maitzen (2002) observed NGC 3960 among other clusters in their
search for chemically peculiar stars using the $\Delta$a system (a three filter
photometric narrow band system), and found only one. 
Finally, Mermilliod et al. (2001) determined precise radial velocities for 
14 red giants, and UBV photoelectric magnitudes for 6 of them.

NGC 3960 has been included in two recent papers aimed at analyzing open
clusters properties in a homogeneous way:
Twarog et al. (1997), on the basis of J81 data, found a 
Galactocentric distance R$_{\rm GC} $ = 7.95 kpc, a distance from the Sun of
1.73 kpc, and a metallicity [Fe/H]=--0.17;
Carraro, Ng \& Portinari (1998) applied the synthetic CMD method,  using J81
photometry, Friel \& Janes (1993) metal abundance, and the Padova tracks
(Bertelli et al. 1994), determining R$_{\rm GC} $ =  8 kpc, and an age of 0.6
Gyr. 

 We describe our photometric data in Section 2; CMDs resulting from the
photometry are presented in Section 3, while Section 4 is devoted to their
application to derive the cluster parameters. Spectroscopic analysis and 
derivation of the cluster chemical abundances are presented in Sections 5 and
6. Finally, results are discussed and  summarized in Section 7.

\section{Photometric Observations and data reduction}

NGC 3960 was observed with DFOSC (Danish Faint Object Spectrograph and Camera)
at the 1.54m Danish telescope located in La Silla, Chile, on 2001 January
24, and 2001 May 16 and 17. 
DFOSC was equipped with the back-illuminated CCD EEV 42-80 (2048x4096
pixels, but only half of the CCD is actually used) with a scale of 0.39
arcsec/pix and a field of view  of 13.3 $\times$ 13.3 arcmin$^2$. 
We used the Johnson-Bessel-Gunn $B$, $V$, $i$
(ESO 450, 451, and 425) and $U$ (ESO 632, in May only) filters;
Table 1 gives a log of the observations.
Two positions were observed, one centered on the cluster, and a second about
30 arcmin away, to be used as comparison for field stars contamination;
Fig.~\ref{fig-map} shows our pointings.

All frames were trimmed and corrected for bias and flat fields in the 
standard way, using IRAF\footnote{
IRAF is distributed by the National Optical Astronomical Observatories, which 
are operated by the Association of Universities for Research in Astronomy, 
under contract with the National Science Foundation} tasks. 
We then used  DAOPHOT--II, also in IRAF environment, to find and measure stars
(Stetson 1987, Davis 1994). All frames were searched independently, using the
appropriate value for the FWHM of the stellar profile and a threshold of 4
$\sigma$ over the local sky value. Instrumental magnitudes were
measured with a Moffat PSF. 
The resulting catalogues were selected in error ($\sigma_{\rm DAO} \le$
0.1 mag), in $\chi^2$ (only to eliminate the very bad cases, i.e. the 
almost saturated stars), and in sharpness (a shape defining parameter, useful
to discriminate between stars and spurious or extended objects). 
The output catalogues were then aligned to a reference frame in each filter, 
for which aperture corrections (i.e., the difference between aperture and PSF
magnitudes) and extinction corrections were derived. Finally, the 
instrumental $u$, $b$, $v$, and $i$ magnitudes for each star were computed  from
a (weighted) average of the individual values.
Using dedicated software by P. Montegriffo (private communication), all frames 
were aligned to the same coordinate system, then astrometrized to the GSC2.

\begin{table*}
\begin{center}
\caption{Observing log, with exposure times in seconds for each filter, and seeing
excursions}
\begin{tabular}{ccccccccc}
\hline
Field & RA(2000) &DEC(2000) & UT Date & U & B & V & I & seeing\\  
\hline
Central & 11:50:42.5 &-55:40:33 &January 24 2001 
        & & 480,80,20 & 240,40,10 & 240,40,10 & 0.94 - 1.2\\
        &            &          &May 16, 17 2001 
	& 900,900,60 &    5 &    5 &      1 & 1.1 - 1.5\\
External &11:54:19.3 &-55:41:06 &May 17 2001 
        & & 600,60,5 & 600,60,1 & 600,60,1 & 1.2 - 1.6\\
\hline
\end{tabular}
\end{center}
\label{tab-log}
\end{table*}

\begin{table}
\begin{center}
\caption{Completeness ratios (cB, cV, cI) in the three bands for the central (cols. 2 to 4)
and external (cols. 5-7) field.}
\vspace{5mm}
\begin{tabular}{rcccccc}
\hline
   mag  &cB &  cV&  cI & cB &  cV&  cI  \\
        &\multicolumn{3}{c}{Centre} 
	&\multicolumn{3}{c}{External} \\
\hline
$<$14.50 & 1.0	 & 1.0   &  1.0     &  1.0  &	1.0   &  1.0   \\
   14.50 & 1.0	 & 1.0   &  0.98    &  1.0  &	1.0   &  1.0   \\
   15.00 & 1.0	 &  0.99 &   0.94   &  1.0  &	0.99  &  1.0   \\
   15.50 & 1.0	 &  0.98 &   0.95   &  0.99 &	0.98  &  1.0   \\
   16.00 & 1.0	 &  0.93 &   0.94   &  0.97 &	0.95  &  0.98  \\
   16.50 & 0.99  &  0.97 &   0.95   &  0.96 &   0.94  &  0.96  \\
   17.00 & 0.98  &  0.95 &   0.92   &  0.95 &   0.93  &  0.93  \\
   17.50 & 0.95  &  0.93 &   0.92   &  0.96 &   0.97  &  0.94  \\
   18.00 & 0.88  &  0.92 &   0.90   &  0.92 &   0.90  &  0.93  \\
   18.50 & 0.92  &  0.90 &   0.87   &  0.91 &   0.88  &  0.91  \\
   19.00 & 0.92  &  0.88 &   0.86   &  0.93 &   0.86  &  0.89  \\
   19.50 & 0.90  &  0.90 &   0.84   &  0.86 &   0.89  &  0.89  \\
   20.00 & 0.89  &  0.88 &   0.79   &  0.89 &   0.85  &  0.86  \\
   20.50 & 0.88  &  0.87 &   0.65   &  0.84 &   0.82  &  0.83  \\
   21.00 & 0.86  &  0.84 &   0.30   &  0.80 &   0.77  &  0.74  \\
   21.50 & 0.85  &  0.77 &   0.00   &  0.74 &   0.71  &  0.21  \\
   22.00 & 0.77  &  0.64 &   0.00   &  0.35 &   0.59  &  0.00  \\
   22.50 & 0.66  &  0.11 &   0.00   &  0.00 &   0.07  &  0.00  \\
   23.00 & 0.20  &  0.00 &   0.00   &  0.00 &   0.00  &  0.00  \\
   23.50 & 0.00  &  0.00 &   0.00   &	    &	      &	       \\
\hline
\end{tabular}
\end{center}
\end{table}

\begin{figure*}
\includegraphics[scale=.8]{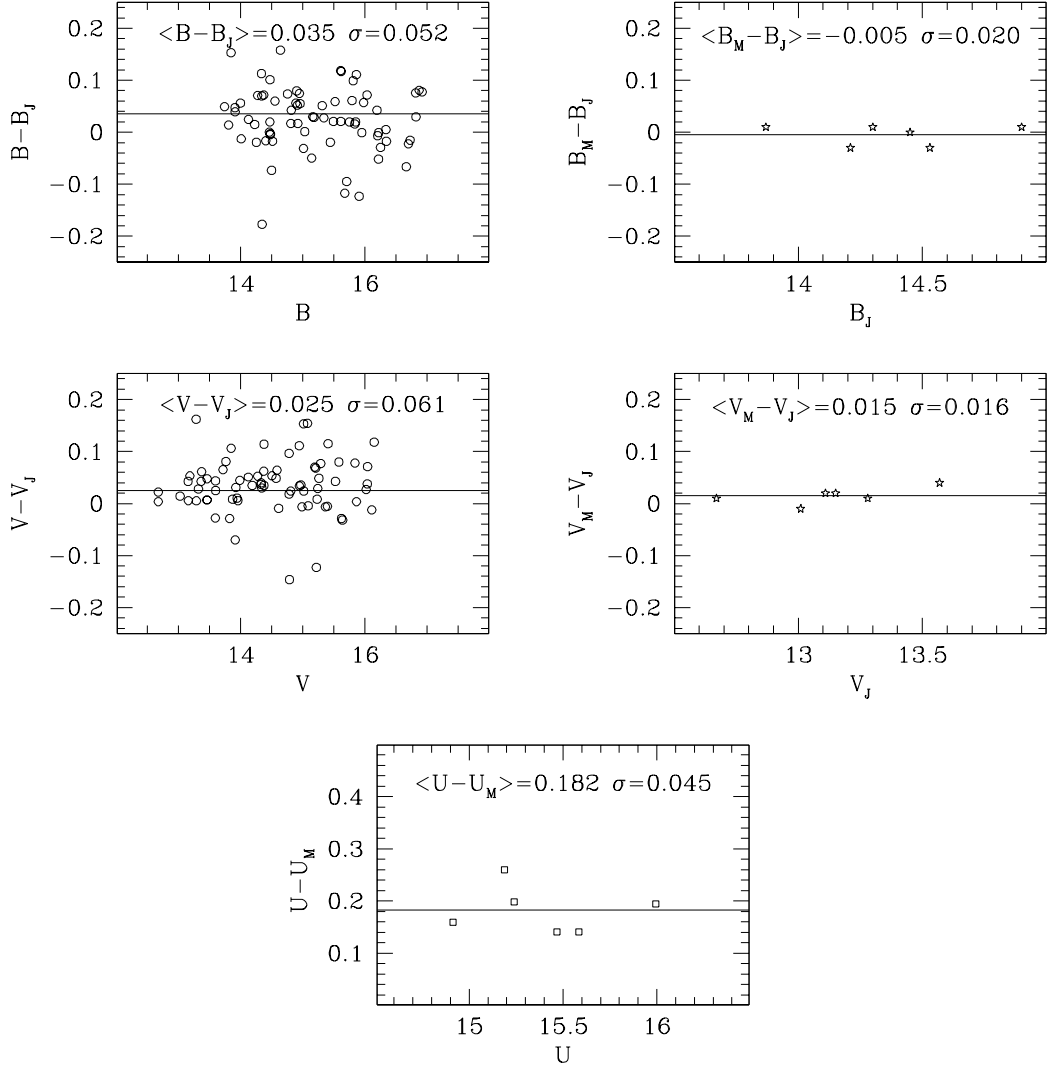}
\caption{Comparison between our photometry and the photolectric one by
J81 (for the B and V filters, 74 stars) or Mermilliod et al. (2001) 
for the U band (6 stars).
The 2 panels on the right show the differences in B and V photolectric
measures for the 6 stars in common.}
\label{fig-conf}
\end{figure*}

\subsection{Calibration to the standard system}

The January 2001 night was photometric with good seeing conditions;
the main observing program of that night was on RR Lyrae stars in 
the Large Magellanic
Cloud, for which excellent calibrations and a precise determination of
the night extinction were essential. Complete description of the
procedure can be found in Clementini et al. (2003) and Di Fabrizio et al. 
(2005), from which we adopt the calibration.
The extinction coefficients for the night were found to be
$K_B = 0.240$, $K_V = 0.142$,  and $K_i = 0.071$, which well agree with 
the average ones derived for La Silla. 27 stars in   
the Landolt (1992) - Stetson (2000)
standard stars fields PG0918+029, PG0231+051, PG1047+003, and SA98 were
used, 
with $-0.273 < B-V < 1.936$ and $-0.304< V-I < 2.142$,  and 
the following  calibration equations were derived:
\[  B-b =  0.1106 \times (b-v) - 0.472 ~~~(r.m.s.= 0.032) \]
\[  V-v =  0.0213 \times (b-v) - 0.175 ~~~(r.m.s.= 0.017) \]
\[  I-i = -0.0227 \times (v-i) - 1.507 ~~~(r.m.s.= 0.025) \]
where $B, V, I$ are in the Johnson-Cousins system, while $b$, $v$, $i$ are the
instrumental magnitudes. These equations, that will be used as 'master
calibrations' for our data, have been computed putting together standards
observed in both nights of the run.

For the May 2001 run we derived new calibrations, since we also have $U$ band
data, and a second, separate, field. We computed aperture photometry for 9
stars in the  two areas PG1323--086 and PG1657+078 (with $-0.149 < B-V < 1.069$
and  $-0.127< V-I < 1.113$), and used average extinction coefficients 
for La Silla ($K_U = 0.44 $, $K_B = 0.21$, $K_V = 0.13$,  and $K_i =
0.06$) to derive the following equations:
\[  U-u =  0.0955 \times (u-b) - 2.9086 ~~~(r.m.s.= 0.028)\]
\[  B-b =  0.1108 \times (b-v) - 0.4674 ~~~(r.m.s.= 0.020)\]
\[  V-v =  0.0103 \times (b-v) - 0.1673 ~~~(r.m.s.= 0.011)\]
\[  I-i = -0.0177 \times (v-i) - 1.5368 ~~~(r.m.s.= 0.020)\]
These equations give the same $B$ and $V$ of the previous ones (to about a few
thousandths of magnitude), and a very similar $I$ (to about 0.02 mag). The ones
involving $B, ~V, ~I$ will be used only to independently calibrate  the
external field. The transformation for the $U$ band can not be compared the
same way.

We have further checked our transformations with the
photoelectric measures existing for some of our stars; we used J81
for the $B$ and $V$ filters,
and Mermilliod et al. (2001) for the $U$ band\footnote{
Note that the two photolectric photometries are quite consistent with each 
other (see right-hand panels of Fig.~\ref{fig-conf})}.
Stars were identified, and we found 74 and 6 objects in common, respectively.
Differences between our photometry and the photoelectric measures are shown in
Fig.~\ref{fig-conf}: on average they are 0.035 and 0.025 mag in the $B$ and 
$V$ bands, respectively.
The case for $U$ is much worse, with a difference of 0.182 mag; a likely
explanation is the use of average extinction coefficients instead of 
specific
ones (unknown to us), that would impact
mostly on the $U$ band. 
Given the general better quality of photoelectric measurements, we shifted
our data in $U$, $B$ and $V$ according to the previous comparison. 
The final calibrated catalogue for NGC 3960
will be accessible through the BDA\footnote{presently at 
{\em http://www.univie.ac.at/webda/new.html}} (Mermilliod 1995). 
 
\begin{figure}
\includegraphics[bb=30 180 430 630, clip, scale=0.6]{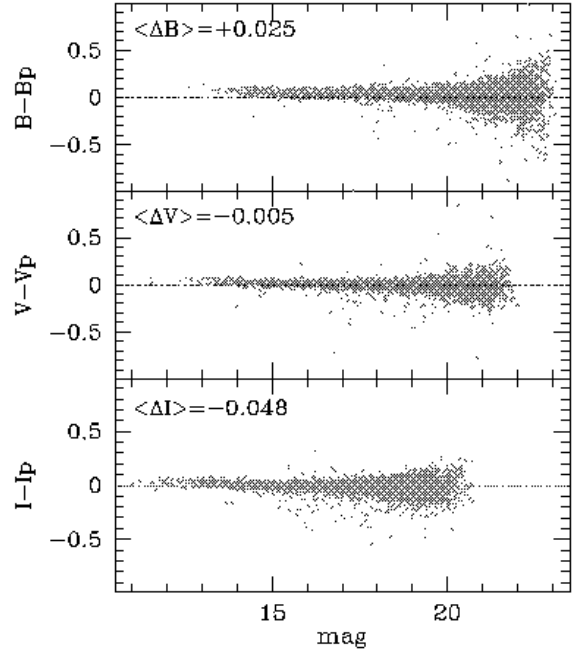}
\caption{Difference in $B$, $V$, $I$ magnitudes between our photometry and P04.}
\label{fig-pris}
\end{figure}

Our data cover the central part of the WFI field studied by P94; we have cross
identified our objects with the ones in their original catalogue  (i.e.,  before
their correction for differential reddening, kindly made available by L.
Prisinzano). We have about 5700 stars in common, and results of the comparison are
shown in Fig. \ref{fig-pris}: the average difference is of +0.025 mag, -0.005 mag,
-0.048 mag in $B$, $V$, $I$ respectively, with a small trend with magnitude. 
The effect
of these differences on the CMDs is negligible, as demonstrated by the very
similar results obtained for the cluster parameters (Sect. 4).

The completeness of our stellar detections was assessed adding  artificial 
stars (50000, at random positions  and selected in magnitude according to the
observed luminosity  function) to the deepest $B$,  $V$, and $I$ images and
exactly repeating the procedure of extraction of objects and PSF fitting used
for the original frame (see e.g. Bragaglia \& Tosi 2003 for more details).  The
output catalogue of the added stars was selected exactly as the science one,
using  error, $\chi^2$  and sharpness.  A star is considered recovered if its
output coordinates coincide (within a fraction of pixel) with the input ones,
and if the difference in magnitude is less than 0.75 mag.  The completeness
level of our photometry at each magnitude is given by the ratio of the number
of recovered artificial stars to the number of added ones and is shown in Table
2.  

\begin{figure*}
\includegraphics[bb=30 230 570 620, clip,scale= 0.8]{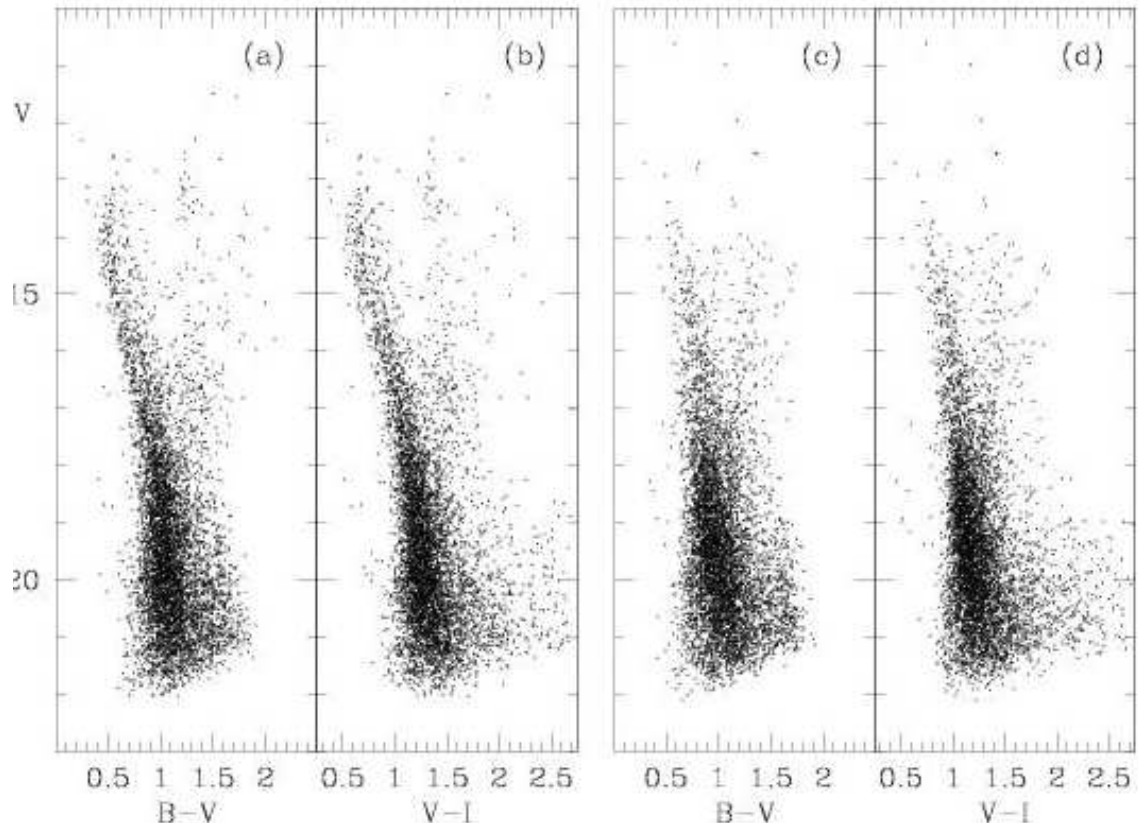} 
\caption{CMDs for the field centered on NGC 3960 
(left-hand panels) and the external comparison one (right-hand panels).}
\label{fig-cmd}
\end{figure*}

\begin{figure}
\begin{center}
\includegraphics[scale=0.9]{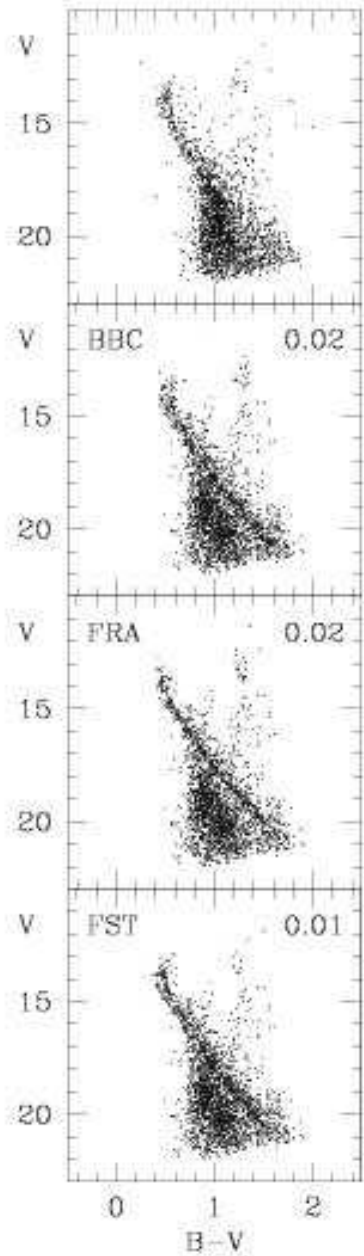}
\caption{Synthetic CMDs in better agreement with the data of the cluster central
region of 4 arcmin radius. 
The empirical CMD is in the top panel. The type of stellar models adopted 
for the synthetic CMDs and their metallicity are indicated in each panel.
See the text for the age, reddening and distance modulus of the
shown models.}
\end{center}
\label{synthcmdbv}
\end{figure}

\begin{figure}
\begin{center}
\includegraphics[scale=0.9]{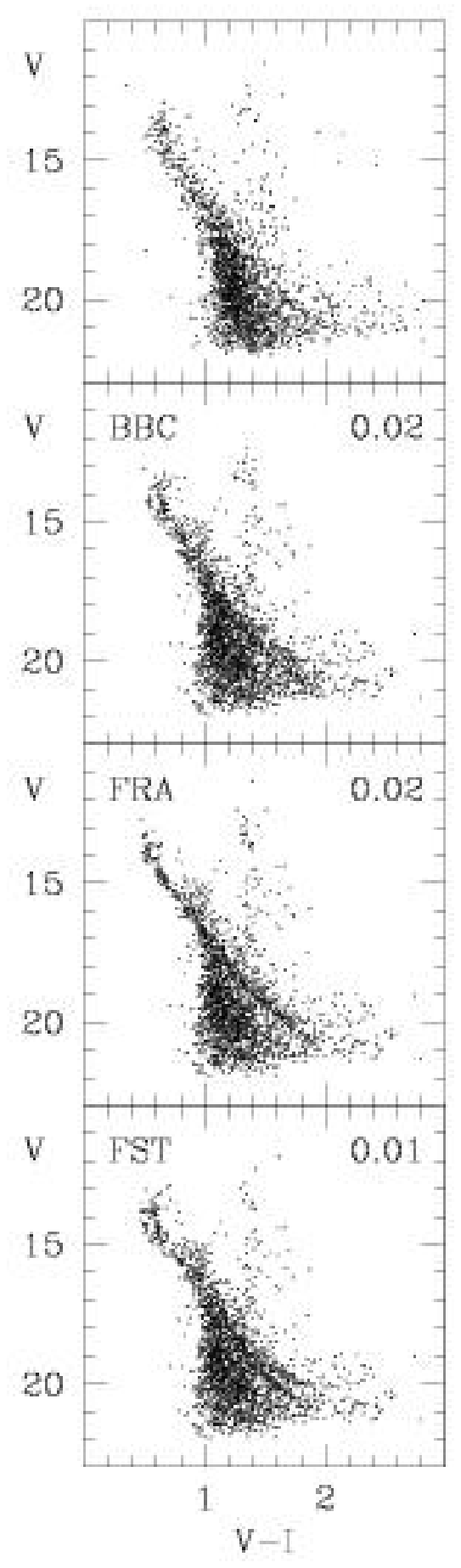}
\caption{V, V--I synthetic CMDs of the models shown in Fig. 6.
The empirical CMD is in the top panel. The type of stellar models adopted 
for the synthetic CMDs and their metallicity are indicated in each panel.}
\label{synthcmdvi}
\end{center}
\end{figure}

\section{The colour-magnitude diagrams}

The final catalogue for the cluster contains 6019 stars with $B, V, I$
magnitudes; the resulting CMDs in $V,B-V$ and $V,V-I$ are shown in Fig.
\ref{fig-cmd}(a,b), while the ones for the external comparison field are in 
Fig. \ref{fig-cmd}(c,d).  

In spite of the relatively large distance ($\sim$30 arcmin) from the cluster
centre of the observed  external field, its CMD has a MS similar to that of 
the cluster, suggesting that stars formed in NGC 3960 are spread over a large
area, as often found for open clusters having suffered strong evaporation. We
have plotted the CMDs of regions with increasing distance from the  cluster
centre and found that beyond a radius of $\sim$4 arcmin the field 
contamination and the spread of the various evolutionary sequences increase
significantly, while the number of probable cluster members (such as clump
stars) remains roughly unchanged.  We will thus restrict the analysis with
synthetic CMDs to the central region  of 4 arcmin radius.

The reddening maps by Schlegel et al. (1998) give a value of \ebv = 0.424
mag for the cluster centre position, but this appears to be an overestimate.
Literature determinations all converge to \ebv = 0.29, and our own analyses,
both photometric (Section 4) and spectroscopic (Section 5), confirm it. 
Furthermore, as noted by the referee, NGC~3960 is at a distance about two thirds
of that of the edge of the dust disc in its direction; in the not unreasonable
assumption than the dust is approximately evenly distributed, only two thirds
of the dust would be in front of the cluster and
its reddening value would be justified.

Given this concordance, we have used the two-colours diagram to confirm the
validity of our $U$ magnitudes rather than to derive a reddening value. We have
considered the location of stars of the well studied OC NGC 752, with [Fe/H] =
$-0.09$ (i.e., very similar to the one we derive for NGC 3960, see Section 6),
for which we adopted the photoelectic $UBV$ data by Johnson (1953) and \ebv =
0.034 (data and information are taken from the BDA). With \ebv = 0.29, its
sequence matches the one of  NGC 3960 reasonably well, and the fit is even
better if our $U$ values are made 0.03 mag fainter, a confirmation that our $U$
photometry is quite well - but not perfectly- calibrated. We may reasonably
assume  this as the residual uncertainty affecting our U data calibration.

\begin{table*}
\begin{center}
\caption{Log of the spectroscopic observations and information on the three 
observed stars.}
\begin{tabular}{cccccccccccc}
\hline
Id  & Id   &  date-obs  & exptime &  Ra(2000)  & Dec(2000)  &  B    &  V    &  I    &  J    &  K      \\
BDA &      &            & (s)     &  (deg)     &  (deg)     &	    &	    &	    & 2MASS &2MASS   \\
\hline 
28  & 4358 & 2001-02-22   & 2$\times$3600    &177.611607 &-55.674498 &14.190 &12.999 &11.774 &10.815 &10.254 \\
    &      & 2001-04-25   & 3600             &&&&&&&& \\       
    &      & 2001-04-26   & 3600             &&&&&&&& \\       
41  & 4351 & 2001-04-26   & 3$\times$3600    &177.650183 &-55.701565 &14.435 &13.131 &11.757 &10.671 &10.080 \\
50  & 4324 & 2001-04-25   & 2$\times$3600    &177.656742 &-55.670876 &14.337 &13.128 &11.810 &10.820 &10.295  \\
\hline
\end{tabular}
\label{t:spe1}
\end{center}
\end{table*}

\begin{table*}
\begin{center}
\caption{Atmospheric parameters and iron abundances for the three clump stars.}
\begin{tabular}{cccccccccccc}
\hline
Id  &Id   & $T_{eff}$ & $\log g$ & [A/H] & $v_t$       & nr & $\log \epsilon$ & $\sigma$ 
                                                       & nr & $\log \epsilon$ & $\sigma$ \\
BDA &	  &    (K)    &          &       & kms$^{-1}$  & Fe {\sc i}& Fe {\sc i}& Fe {\sc i}& Fe {\sc ii}& Fe {\sc ii}& Fe {\sc ii}\\
\hline
28  &4358 & 4900& 2.06& -0.15& 1.23& 98 & 7.402 & 0.108 & 13 & 7.346 & 0.190  \\
41  &4351 & 4850& 2.20& -0.14& 1.21& 97 & 7.390 & 0.132 & 14 & 7.335 & 0.176  \\
50  &4324 & 5000& 2.70& -0.06& 1.15&106 & 7.478 & 0.112 & 12 & 7.431 & 0.182 \\
\hline
\end{tabular}
\label{t:spe2}
\end{center}
\end{table*}

\section{Derivation of the cluster parameters from the photometry}

We have derived age, reddening and distance modulus of NGC 3960 from the
photometric data, using the
synthetic CMD method (Tosi et al. 1991), as for all the clusters of our project
(see Kalirai \& Tosi 2004; Bragaglia \& Tosi 2005).
As usual, we have applied the method adopting three different sets of stellar
evolution models, in order to estimate the theoretical uncertainty on the
resulting values. The adopted evolutionary tracks are those computed by the 
Padova group (Bressan et al. 1993, Fagotto et al. 1994; hereafter BBC) 
considering overshooting from convective regions, 
the Full Spectrum Turbulence models computed by Ventura et al. (1998; 
hereafter FST), with and without overshooting, 
and the models with semiconvection and no overshooting computed by the 
Frascati group (Dominguez et al. 1999; hereafter FRA). 
More details on the method and a summary of the
results obtained for all the clusters studied so far are provided by
Bragaglia \& Tosi (2005).

The synthetic CMDs have been created assuming a Salpeter Initial Mass Function,
an (almost) instantaneous burst of star formation,
the photometric errors and the
completeness factors derived from the extensive artificial star tests on the 
data images 
described above. Theoretical luminosity and effective temperature of the
synthetic stars have been transformed to magnitudes and colours using Bessel,
Castelli \& Pletz (1998) photometric conversions to the Johnson-Cousins system.

For the reasons presented in the previous Section, we have decided
to consider only a region within a 4 arcmin radius of the cluster centre, 
that contains 1966 stars. Since equal area zones of the external field 
contain between 1750 and 1840 stars, if none of these were attributable to
NGC 3960, the cluster members should be at most 216. However, synthetic CMDs
with this total number of cluster objects grossly underproduce the 
number of stars on the
upper MS and post-MS phases. We therefore believe that many low-mass stars
have evaporated from the cluster. We have iteratively considered as
representative of the original number of objects formed in the cluster only the
number of existing stars brighter than a given magnitude (i.e. more massive than
a given mass). It turns out a posteriori that only stars brighter than V=15
appear to be unaffected by evaporation. In fact, synthetic models with 90 
stars\footnote{90 = 111 - 21 being the number of stars with
V$\leq$15 in  the central region of 4 arcmin radius minus the
corresponding number in the external field}
brighter than V=15  reproduce very well the empirical luminosity function. 
Such models predict between 400 and 440 stars still alive in NGC 3960, about
twice as many as the 216 obtained subtracting the number of stars observed in
the control field.  This indicates the high level of
evaporation occurred during the cluster lifetime (much longer than the 
cluster relaxation time).
 
The synthetic CMDs allow for a fraction of unresolved binary systems from 0
to 50\%, assuming a random mass ratio between the primary and the secondary star
of the system. Given the large colour spread in the sequences of the
observational CMD of NGC 3960, varying amounts of differential reddening have
also been considered. We have found that to simultaneously reproduce the 
shape of the TO and of the MS, their colour spread and distribution, on
average we need to assume with all the stellar models a fraction of binaries 
of 50\% and a differential reddening $\Delta$\ebv ~of about 0.1. Models with 
smaller differential reddening have tighter MS and may be acceptable with the 
BBC tracks but not with the FRA ones. Models with less binaries and larger 
$\Delta$\ebv ~(or vice versa) do not reproduce equally well the observed 
features, since the two effects widen the MS toward the red in different ways. 

For sake of homogeneity with the approach followed for the other clusters of the
project for which no metallicity estimate from high-resolution spectroscopy is
available, we have run the simulations for various metallicities as if we
had not derived the cluster abundances from our own spectra. The comparison of
the model metallicity leading to synthetic CMDs in better agreement with the
photometric data with that derived from the spectroscopic ones is then a
check of how well the theoretical abundances compare with the actual ones.
The tested metallicities are: Z=0.004, Z=0.008 and Z=0.02 with the BBC models,
and Z=0.006, Z=0.01 and Z=0.02 with the FRA and FST models.
We find that the synthetic CMDs with Z=0.004, 0.006 and 0.008 all lead to
unsatisfactory results, because they do not reproduce either the shape of the MS
or of the TO, or they overpopulate the clump. They all need reddenings \ebv 
~of at least 0.40, larger than both the literature and the spectroscopic values.

Models with Z=0.01 and 0.02, instead, reproduce quite well the observed
features of both the V, B--V and the V, V--I diagrams. With the FST models, we
find that to reproduce the observed features of NGC 3960, the evolutionary
tracks assuming the maximum overshooting parameter ($\eta$=0.03, see Ventura et
al. 1998) are more appropriate. The best 
agreement with the data is reached with Z=0.01 (with age 0.9 Gyr, \ebv
= 0.30 $\pm \Delta$\ebv, and \mmm = 11.5), although the Z=0.02 model (with age 
0.8 Gyr, \ebv = 0.28 $\pm \Delta$\ebv, and \mmm = 12.0) is not bad either. 
With the FRA models the agreement is less good, because they all overpopulate
the clump and have the TO shape more vertical than observed. It is difficult 
to choose between the Z=0.01 models (with age 0.5 Gyr, \ebv = 0.43 
$\pm \Delta$\ebv, 
and \mmm = 11.3) and the Z=0.02 ones (with age 0.6 Gyr, \ebv = 0.31 
$\pm \Delta$\ebv,
and \mmm = 11.6), but the latter look slightly more adequate.
BBC models are not available for metallicity values between 0.008 and
0.02, and the solar ones reproduce well the data (although not as well as the
FST Z=0.01 ones, because of a slight overpopulation of the clump) with age 0.9 
Gyr, \ebv = 0.27 $\pm \Delta$\ebv, and \mmm = 11.7.
$\Delta$\ebv \ is equal to 0.05  for all these models; it  indicates the
differential reddening and not the error on the average reddening. Notice how
small the spread on the three parameters resulting from the theoretical
uncertainties is: for the best models, the average reddening is between 0.27
and 0.31, the distance modulus between 11.5 and 11.7, and the age is 0.9 Gyr
with both the BBC and FST overshooting models and obviously less, 0.6 Gyr, with
the FRA no-overshooting ones.

\begin{figure}
\begin{center}
\includegraphics[bb=120 140 450 365, clip,scale=.7]{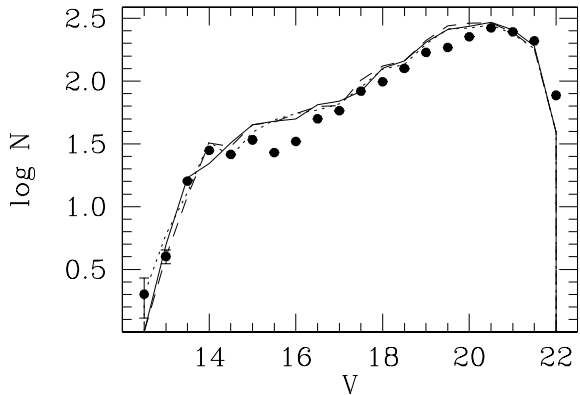}
\caption{Comparison between the cluster luminosity function of the central
4 arcmin radius region (circles) and those (lines) corresponding to the
synthetic CMDs of Figs 6 and 7. The solid line
refers to the BBC models, the dotted one to the FRA models and the dashed line
to the FST models. The error assumed on the observed LF is Poissonian, and 
error bars are always smaller than the points, except for the first two.}
\label{synthlf}\end{center}
\end{figure}

The synthetic CMDs in better agreement with the cluster data for each of the
three types of stellar evolution models are shown in Figs 6 
and 
\ref{synthcmdvi}. In these diagrams, to the synthetic stars we have 
overimposed the 1757 stars 
measured in an equal area zone of the external field. In both figures, the
top panel shows on the same scale the empirical CMD of the central 
4 arcmin radius region, to be directly compared with the synthetic ones.
The corresponding luminosity functions are shown in Fig. \ref{synthlf}, where the
lines refer to the models and the circles to the data.

The striking difference seen in the CMDs between the observed lower MS
population  and the one modeled is due to the strong evaporation of low-mass
cluster stars and is also visible in the LFs. The LFs of the different best
cases coincide at  faint magnitudes, where the LF is dominated by field
objects, and are barely distinguishable from each other at bright magnitudes.

\begin{table}
\begin{center}
\caption{Values for the reddening \ebv \ computed from the spectroscopic temperatures
and the Alonso et al. (1999) calibrations, derived from  the $B-V$  and the $V-K$
colours. Last column gives the average values.}
\begin{tabular}{ccccc}
\hline
Id  &Id   & \ebv    & \ebv     & \ebv \\
BDA &     & ($B-V$) & ($V-K$)  & average \\
\hline
28  &4358 & 0.227   & 0.226    & 0.2265\\
41  &4351 & 0.365   & 0.356    & 0.3605\\
50  &4324 & 0.320   & 0.315    & 0.3175\\
\hline
\end{tabular}
\label{t:red}
\end{center}
\end{table}

\begin{table*}
\begin{center}
\caption[]{Abundances of the elements measured in the three stars of NGC
3960, where n is the number of lines used. The reference Solar value is given
in col. 2.}
\begin{tabular}{lcrrcrrcrrcrcc}
\hline
Element       & Sun &\multicolumn{3}{c}{star 28/4358} &\multicolumn{3}{c}{star 41/4351}
              &\multicolumn{3}{c}{star 50/4324} & mean & $\sigma$ & Note\\
	      &   & n  &  &$\sigma$ & n  &. &$\sigma$ & n  & 
	      &$\sigma$ & & & \\
\hline
${\rm [Fe/H]}${\sc i}  & 7.54  & 98 &$-$0.14 &0.11 & 106 &$-$0.06    &0.11 & 97 &$-$0.15 &0.13 & $-$0.12 & 0.05 & \\
${\rm [Fe/H]}${\sc ii} & 7.49  & 13 &$-$0.14 &0.19 &  12 &$-$0.06    &0.18 & 14 &$-$0.16 &0.18 & $-$0.12 & 0.05 & \\ 
${\rm [Na/Fe]}${\sc i} & 6.21  &  4 &   0.33 &0.11 &   5 &	0.17 &0.12 &  5 &   0.47 &0.10 &    0.32 & 0.12 & N-LTE\\ 
${\rm [Mg/Fe]}${\sc i} & 7.43  &  5 &   0.17 &0.06 &   7 &	0.01 &0.09 &  6 &   0.21 &0.09 &    0.01 & 0.08 & \\ 
${\rm [Al/Fe]}${\sc i} & 6.23  &  4 &   0.02 &0.13 &   5 &$-$0.09    &0.12 &  4 &   0.10 &0.16 &    0.13 & 0.09 & \\
${\rm [Si/Fe]}${\sc i} & 7.53  & 20 &   0.23 &0.21 &  17 &	0.19 &0.15 & 19 &   0.27 &0.21 &    0.23 & 0.03 & \\
${\rm [Ca/Fe]}${\sc i} & 6.27  & 19 &   0.16 &0.18 &  17 &	0.04 &0.17 & 18 &   0.16 &0.20 &    0.12 & 0.06 & \\
${\rm [Sc/Fe]}${\sc ii}& 3.13  &  7 &   0.02 &0.19 &   5 &	0.19 &0.25 &  7 &$-$0.06 &0.20 &    0.05 & 0.10 & HFS\\
${\rm [Ti/Fe]}${\sc i} & 5.00  & 30 &$-$0.04 &0.16 &  29 &$-$0.12    &0.21 & 30 &$-$0.10 &0.15 & $-$0.09 & 0.04 & \\
${\rm [Ti/Fe]}${\sc ii}& 5.07  & 14 &$-$0.07 &0.12 &  10 &	0.02 &0.11 &  9 &$-$0.13 &0.18 & $-$0.06 & 0.06 & \\
${\rm [V/Fe]}${\sc i}  & 3.97  &  9 &$-$0.02 &0.16 &   9 &$-$0.10    &0.16 &  8 &$-$0.05 &0.18 & $-$0.06 & 0.03 & HFS\\ 
${\rm [Cr/Fe]}${\sc i} & 5.67  & 36 &   0.06 &0.22 &  43 &$-$0.02    &0.18 & 35 &   0.03 &0.23 &    0.02 & 0.03 & \\
${\rm [Cr/Fe]}${\sc ii}& 5.71  &  7 &   0.01 &0.15 &  11 &	0.01 &0.11 & 12 &$-$0.08 &0.21 & $-$0.02 & 0.04 & \\
${\rm [Mn/Fe]}${\sc i} & 5.34  &  6 &   0.12 &0.19 &   6 &$-$0.11    &0.18 &  3 &$-$0.06 &0.22 & $-$0.02 & 0.10 & HFS \\
${\rm [Co/Fe]}${\sc i} & 4.29  &  5 &$-$0.06 &0.09 &   2 &$-$0.04    &0.03 &  4 &$-$0.08 &0.22 & $-$0.06 & 0.02 & HFS \\
${\rm [Ni/Fe]}${\sc i} & 6.28  & 34 &$-$0.06 &0.11 &  40 &$-$0.08    &0.21 & 36 &$-$0.01 &0.17 & $-$0.05 & 0.03 & \\
${\rm [Ba/Fe]}${\sc ii}& 2.22  &  3 &   0.55 &0.08 &   3 &	0.49 &0.42 &  3 &   0.62 &0.11 &    0.55 & 0.06 & \\ 
\hline
\end{tabular}
\label{t:abutot}
\end{center}
\end{table*}

\section{Spectroscopic data}

Three clump stars in NGC 3960 were selected among the true cluster members on
the basis of the radial velocities (RVs) measured by Friel \& Janes (1993),
later confirmed by Mermilliod et al. (2001) and by our own measures.   
They were observed in 2001 (February and April) with FEROS (Fiber-fed Extended
Range Optical Spectrograph) mounted at the 1.5m telescope in La Silla (Chile)
at R = 48000, with a wavelength coverage of $\lambda\lambda$ 3700-8600 \AA; a
log of the observations is given in Table \ref{t:spe1}, together with
photometric information. The averaged spectra for each stars have high signal
to noise ratio; S/N is 100 for stars 28 and 41, and 80 for star 50 (measured
near 6150 \AA).  The heliocentric RVs perfectly agree with the ones in
Mermilliod et al. (2001) for stars 28 and 41 (4358 and 4351 in our numbering
system). Star 50 (4324 in our numbering system) is a long period spectroscopic
binary and our 3 spectra  taken on the same night can not produce a significant
average value; with RV $\sim$ -15  km s$^{-1}$ they appear compatible with the
known RV and amplitude of the RV curve (Mermilliod et al. 2001: RV = $-$22.31
km s$^{-1}$, K = 12.92 km s$^{-1}$). 

These spectra have been obtained, reduced, and analyzed exactly as described in
Bragaglia et al. (2001) and Carretta et al. (2004, 2005). Here  we present 
the results and refer to those papers for a detailed description of the method
used. 

Equivalent widths ($EW$s) were measured employing an updated version of  the
ROSA spectrum analysis package (Gratton
1988). As usual, we restricted to the 5500-7000~\AA\ spectral range 
for Fe lines, and employed the entire spectrum for the other species.
Sources of oscillator strengths and atomic parameters are the same as Gratton
et al. (2003).

Effective temperature ($T_{eff}$) and gravity ($\log g$) for each star were
derived from the spectra using the excitation and ionization
equilibria for iron, respectively. The microturbulent velocity ($v_t$) was derived
assuming the relation between $\log g$ and $v_t$ given in Carretta et al. (2004).
These parameters, along with iron abundances, are shown in Table \ref{t:spe2}.
Errors were estimated as in Carretta et al. (2004).
They comprise a systematic part (due e.g., to uncertainties in the
adopted oscillator strengths) and a random part (due e.g., to the different
S/N ratios) that represents the internal error; 
random errors have been found to be 62 K in $T_{eff}$,
0.25 dex in $\log g$, 0.17 km s$^{-1}$ in $v_t$, and 0.05 dex in [A/H].
Sensitivity of the parameters on the internal errors can be found in tab. 3
of Carretta et al. (2005).

As our spectroscopically determined temperatures are
reddening-free, we can derive an independent measure of the reddening.
We entered the spectroscopically determined $T_{eff}$'s into 
the colour-temperature transformations by Alonso, Arribas \&  
Mart{\'{\i}}nez-Roger  (1999) and
obtained de-reddened colours to be compared to the observed ones based on
our $B, V$ photometry and the 2MASS $J, K$ values (Cutri et al. 2003).
The reddening for the three stars is given in Table \ref{t:red}. We estimate
errors on these \ebv's to be at most 0.02-0.03 mag, assuming a conservative
error of $\pm$ 0.1 dex in metallicity and 90 K in $T_{eff}$. \ebv \ values from
$B-V$ and $V-K$ [adopting $E(V-K)=2.75 E(B-V)$; Cardelli et al. 1989] are almost
identical for each star (less than 0.01 mag at the most).  The average
reddening is \ebv = 0.301, with large differences among  the three stars (rms =
0.068). Even if the spectroscopic sample is very  small, both 
average value and scatter around it are
compatible with what we derived (\ebv = 0.29, $\Delta$\ebv = 0.05) from the
photometric data of the  central part of the cluster, where all three stars are
located.

\section{Chemical Abundances}

The iron abundances derived from $EW$s were checked using synthetic spectra of
about 20-25 selected iron lines,  as amply described in Carretta et al. (2004). The
average difference between abundances derived from  $EW$s and from synthesis 
is $-0.05$ dex,
without a systematic trend, so we deemed to have achieved the accuracy
in continuum tracing and $EW$ measurement attainable with these spectra.   
When we average the three stars we obtain [Fe/H] = $-$0.12 $\pm$ 
0.02, rms 0.05 dex; this error bar is simply the standard error of the mean.

This metallicity is based on high resolution
spectroscopy, the best technique to derive accurate chemical abundances, so it
should supersede other measurements, to which anyway we may compare it.
From the synthetic CMD method we find Z
= 0.01 or 0.02, i.e. [Fe/H]=$-$0.3 or 0.  Literature values (see
Introduction) range from [Fe/H] = $-$0.68 (based on Washington
photometry, that gives systematically lower results) to $\simeq -0.30$ (DDO
photometry and low resolution spectroscopy) or $-$0.06 (another calibration of
DDO photometry). 

Abundances were also derived for the light elements Na, Al, 
for the $\alpha$-process elements Mg, Al, Ca, Ti {\sc i} and
{\sc ii}, the Fe-group elements Sc {\sc ii}, V, Cr {\sc i} and {\sc ii}, Mn, Co, Ni,
and for the $n$-capture element Ba {\sc ii}. 
We corrected the Na abundance for departures from LTE following Gratton et
al. (1999), while those of Sc, V, Mn, Co, and Ba have hyper-fine structure (HFS)
taken into account whenever necessary. Abundances of C, N, O are deferred to a
dedicated paper since they require synthetic spectrum analysis. 
Results for each star and the cluster averages are presented in Table
\ref{t:abutot}, together with the reference solar values adopted.

Abundances from different ionization stages (Ti {\sc i} and Ti {\sc ii},  
Cr {\sc i} and Cr {\sc ii}) are in very good accord,
endorsing the adopted atmospheric parameters. The three stars do not seem
to show scatter in any of the abundances, as we already found for the
other OCs studied. The $\alpha$-elements have a value slightly over solar
(+0.10), while Na is clearly overabundant (+0.32), in line with the old OC
population (Friel et al. 2003, Carretta et al. 2005). NGC 3960 does not show any
striking peculiarity, and its composition will be considered in conjunction with
the ones of the other OCs in our sample in forthcoming papers.

\section{Summary and Discussion}

We have analyzed photometric and high resolution spectroscopic data for the
open cluster NGC 3960, and found age, distance, reddening and metallicity 
using  the synthetic CMD method and fine abundance analysis. Our best estimates
are:   age of  0.9 Gyr or 0.6 Gyr (using stellar models with/without
overshooting),  \mmm = 11.6 $\pm$ 0.1, \ebv = 0.29 $\pm$ 0.02 (with a
differential reddening $\Delta$\ebv = 0.05), metallicity between solar and half
of solar from the photometry and [Fe/H] = $-$0.12 (rms 0.05) dex from the
spectra.

These value are in good agreement with J81 and the most recent investigation of this
cluster (P04), taking into account the different methods and assumptions.

At the distance of NGC 3960 derived from the synthetic CMDs (2.1 kpc from the sun) 
the external field
observed at 30 arcmin from the cluster centre is at 18.3 pc. The fact that it
clearly contains stars falling on the MS of the cluster CMD shows that these
stars have been able to travel this distance from their original birthplace. We
have found that about half of the cluster stars fainter than $V$ = 15 (i.e.
with mass lower than $\sim$1.6 \msun) are likely to have evaporated from the 
central region. 

Existence of segregation and evaporation is in agreement with what P04 found 
from their analysis. They derived the luminosity and mass functions, corrected
for the presence of field stars, for the central part of the cluster (see their
figs. 15 and 16); their mass function drops below about 1 M$_\odot$, because of
the combination of incompleteness and mass segregation that moves the lighter
stars outside the considered radius (larger than ours: 7 arcmin).

The differential reddening values of 0.1 in \ebv ~in the central area of 4 arcmin
radius agrees with the amount proposed by P04 for the same
zone, although they suggest the higher value of $\Delta$\evi=0.57 over their
much larger field of view. The average (central) reddening 
resulting from the best fitting synthetic CMDs is supported by the
two-colours diagram. It also compares very well with the literature values
(except Schlegel et al. 1998) and with that inferred from the spectroscopic
analysis of the clump stars that show some evidence of differential reddening
too. 

Analysis of FEROS spectra of three clump stars was used to 
derive the detailed chemical composition of the cluster, determining abundance
of iron, $\alpha$-elements and heavier elements. Their behaviour is similar
to what has usually been found for old open clusters. 

The cluster metallicity derived from our high-resolution spectra, 
[Fe/H] = $-$0.12, corresponds to a metal mass fraction Z$\simeq$0.015. 
This value is between those
available for the stellar evolutionary tracks and this explains the ambiguity
in the choice between the Z=0.01 and the Z=0.02 synthetic models. We consider
this result as quite favourable, taking into account that the stellar model
metallicity depends on a series of parameters, such as opacities, photometric
conversions, etc.

\bigskip\noindent
ACKNOWLEDGEMENTS

We acknowledge helpful discussions with L. Prisinzano, who also kindly supplied
her original catalogue. We thank P. Montegriffo, whose programs were
used for the data analysis, and G. Clementini for the calibration equations.   
We thank the referee for useful comments and suggestion about the reddening.
The bulk of the simulation code was originally provided by L.Greggio. 
Finally we acknowledge the use of the valuable BDA database, maintained 
for years by
J.-C. Mermilliod in Geneva and recently moved to Vienna and Dr. Paunzen's
care. This
publication makes use of data products from the Two Micron All Sky Survey,
which is a joint project of the University of Massachusetts and the Infrared
Processing and Analysis Center/California Institute of Technology, funded by
the National Aeronautics and Space Administration and the National Science
Foundation. Partial finantial
support to this project has come from the the Italian MIUR through 
PRIN 2003029437.

\end{document}